\documentclass[a4paper]{jpconf}
\usepackage{graphicx,amsmath,hyperref}
\usepackage[small]{caption}
\begin{document}
\title{Medium-induced multi-photon radiation}

\author{Hao Ma, Carlos A. Salgado}
\address{Departamento de F\'isica de Part\'iculas, Universidade de Santiago de Compostela, E-15782 Santiago de Compostela, Spain}
\author{Konrad Tywoniuk}
\address{Department of Astronomy and Theoretical Physics, Lund University, S\"olvegatan 14A, S-223 62 Lund, Sweden}

\begin{abstract}
We study the spectrum of multi-photon radiation off a fast quark in medium in the BDMPS/ASW approach. We reproduce the medium-induced one-photon radiation spectrum in dipole approximation, and go on to calculate the two-photon radiation in the Moli\`{e}re limit. We find that in this limit the LPM effect holds for medium-induced two-photon ladder emission.
\end{abstract}

\section{Introduction}
To date, the medium-modification of the jet fragmentation function has only been calculated for the single-particle emission spectrum. In QED, the one-photon radiation off an energetic quark/electron propagating a dense, finite-sized medium was calculated as a multiple scattering process in \cite{WiedemannGyulassy}. In the context of QCD, the corresponding gluon radiation process, including multiple gluon interactions, was calculated in \cite{BDMPS, Wiedemann, SW, ASW}. In general, since it does not interact, the photon spectrum is hardly suppressed by formation time effects, in contrast to the gluon case. Thus, while the suppression of high-$p_\perp$ hadrons serves as a probe  \cite{HardProbe}  of the hot and dense medium created in the wake of a heavy-ion collision \cite{PHENIX, PHOBOS, BRAHMS, STAR}, photon production is used as a calibration due to its lack of interaction with the surrounding QCD matter \cite{PhotonExperiment}.

On the other hand, medium-modifications of more exclusive observables, e.g., two-particle emission, are hitherto unknown. Particularly in vacuum, the latter process comprise key features such as the energy and angular ordering of QCD radiation which arise due to interference between emitters. In QED, the lack of ordering of the radiation can be traced back to the fact that the photon does not carry charge. On the contrary, intensive medium-interactions of the quark could induce unprecedented interferences between subsequent emissions which, in principle, could be measured by experiment. Furthermore, the two-photon spectrum is expected to be less involved than the corresponding QCD process due to the sterile nature of the photons, and thus serves also as a first attempt at computing multi-particle radiation in medium.

Following the BDMPS/ASW approach \cite{WiedemannGyulassy,BDMPS, Wiedemann, SW, ASW}, we focus on a highly energetic quark traversing the medium with energy $E_q$, which radiates a soft photon with energy $\omega$ and transverse momentum $k_{\perp}$ ($E_q$ $\gg$ $\omega$ $\gg$ $k_{\perp}$). The medium is modeled as a collection of static scattering centers, given by the gauge field $A^a(x)t^a$ described by a Yukawa potential with the Debye mass serving as the inverse screening length, where $t^a$ is the color matrix in the fundamental representation. Due to the interaction with the medium, the hard quark acquires an eikonal phase which is described by the Wilson line. Additionally, due to the Bloch-Nordsieck structure of the QED radiation vertex, we need to take into account corrections from the Brownian motion of the quark in the transverse plane.

\section{Medium-induced one-photon radiation}
Here we are interested in the leading contribution to the spectrum in the high-energy approximation. In QED, one has to keep both order ${\cal O} ((1/p_+)^0)$ and order ${\cal O} (1/p_+)$ terms in the radiation vertex, where $p_+\simeq\sqrt{2}E_q$ is the quark light cone momentum. Accordingly, we derive the spin non-flip medium-induced one-photon radiation spectrum for a finite medium of length L, as in \cite{WiedemannGyulassy}. 

For the purpose of illustration, let us consider spectrum for a fixed number of medium scatterings in the Bethe-Heitler and factorization limits. In the first case, the $N$ scattering centers are well-separated, thus the radiation off these scattering centers is the sum of $N$ Bethe-Heitler spectra, given by
\begin{equation}
\label{eq:BH}
\frac{{\rm d}^5 I^\text{BH}}{{\rm d}^2 {\bf p}_{f \perp} {\rm d}^2 {\bf k}_{\perp} {\rm d} (\ln x)} \propto \frac{\alpha_{em}}{\pi^3}  \sum_i \frac{x^2 \, {\bf q}_{i \perp}^2}{{\bf k}_{\perp}^2 ({\bf k}_{\perp} - x \, {\bf q}_{i \perp})^2} \,,
\end{equation}
where ${\bf p}_{f\perp}=\sum_i {\bf q}_{i\perp} -{\bf k}_\perp$ and ${\bf k}_{\perp}$ are the transverse momentum of the outgoing quark and emitted photon, respectively, while $x$ is given by $\omega=xE_q$, ${\bf q}_{i\perp}$ is the momentum transfer from the $i$th scattering center, and finally $\alpha_{em}$ is the electromagnetic coupling. In the factorization limit, on the other hand, the scattering centers are not resolvable, thus the resulting spectrum is given by
\begin{equation}
\frac{{\rm d}^5 I^\text{fact}}{{\rm d}^2 {\bf p}_{f \perp} {\rm d}^2 {\bf k}_{\perp} {\rm d} (\ln x)} \propto \frac{\alpha_{em}}{\pi^3} \frac{x^2 \, (\sum_i {\bf q}_{i \perp})^2}{{\bf k}_{\perp}^2 ({\bf k}_{\perp} - x \, \sum_i {\bf q}_{i \perp})^2}.
\end{equation}
When $i=1$, we get the expression of the opacity expansion at first order from both of these two limits. The full medium-induced one-photon radiation spectrum interpolates between the above two limits. This is how one incorporates the Landau-Pomeranchuk-Migdal (LPM) effect \cite{LandauPomeranchuk, Migdal}, which accounts for the fact that the scattering centers in medium interfere destructively during the formation time of an on-shell photon.

The coherent nature of the spectrum becomes most pronounced in the so-called Moli\`ere limit, i.e. i) for soft and small angle photon emission keeping the transverse momentum of the photon much larger than the total transverse momentum transfer from the medium,  ii) in case of large length and iii) small density of the medium at hand. In this approximation, we can perform the sum over all multiple soft scatterings and the spectrum then reads
\begin{equation}
\label{OnePhotonSpectrumMoliere}
\frac{{\rm d}^5 I^\text{Mol}}{{\rm d}^2 {\bf p}_{f \perp} {\rm d}^2 {\bf k}_{\perp} {\rm d} (\ln x)} \approx \frac{\alpha_{em}}{\pi^3} \frac{x^2}{{\bf k}_{\perp}^4} \exp \bigg\{- \frac{(\sum_i {\bf q}_{i \perp})^2}{2 \ n_0 \, C \, L_+} \bigg\}=
\frac{\alpha_{em}}{\pi^3} \frac{x^2}{{\bf k}_{\perp}^4} \exp \bigg\{- \frac{({\bf p}_{f \perp}+{\bf k}_\perp)^2}{2 \ n_0 \, C \, L_+} \bigg\},
\end{equation}
where $n_0$ is the density of the medium, $C$ is the probability of the hard quark scattering and $L_+=\sqrt{2}L$ is the size of the medium. In eq.~(\ref{OnePhotonSpectrumMoliere}), we still find the characteristic  $x^2$ rapidity dependence, accompanied by the $1/{\bf k}_\perp^4$ drop, as for the Bethe-Heitler and factorization spectra in the limit $k_\perp \gg q_\perp$. Yet, where as the two latter vanish in the limit $q_\perp\to0$, the Moli\`ere spectrum peaks at zero momentum transfer. This illustrates the fact that the radiation spectrum for multiple soft scattering retains features which cannot be obtained from a calculation at fixed order in opacity.

\section{Medium-induced multi-photon radiation}
\begin{figure}
\begin{center}
\includegraphics[width=5in]{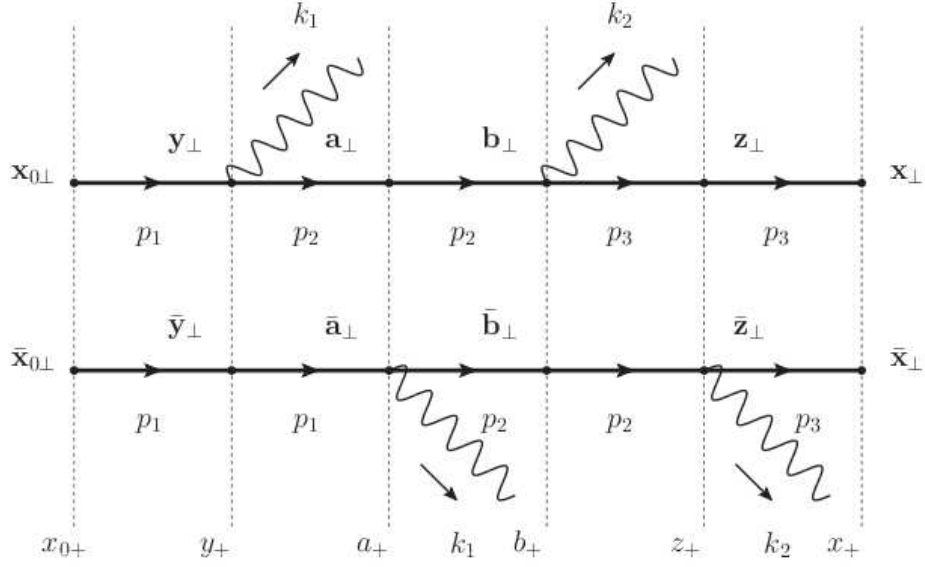}
\caption{\label{twophotonladder1}The radiations are inside the medium in both amplitude and conjugate amplitude. Here $p_1$ $=$ $p_+$, $p_2$ $=$ $(1 - x) \, p_+$ and $p_3$ $=$ $(1 - y) \, (1 - x) \, p_+$.}
\end{center}
\end{figure}

We go on to calculate the radiation of  two photons inside the medium in both the amplitude and the conjugate amplitude for the situation of identical sequence of radiations in the amplitude and conjugate amplitude, respectively. We denote this particular contribution as a {\it ladder} emission. The corresponding diagram is depicted in Figure~\ref{twophotonladder1}. In particular, first a photon is radiated with transverse momentum ${\bf k}_{1 \perp}$ and carrying a momentum fraction $x$ of the parent quark, and subsequently another photon with transverse momentum ${\bf k}_{2 \perp}$ and carrying a momentum fraction $y$ of the quark. In this case, the expression reads
\begin{equation}
\begin{split}
\label{eq:LadderCrossSection}
& \frac{{\rm d}^8 I}{{\rm d}^2 {\bf p}_{f \perp} {\rm d}^2 {\bf k}_{1 \perp} {\rm d}^2 {\bf k}_{2 \perp} {\rm d} (\ln x) {\rm d} (\ln y)} = \frac{(\alpha_{em})^2}{(2 \, \pi)^6} \, \text{Re} \, \frac{2}{x \, y \, (1 - x)^2 \, (p_+)^4} \\
& \quad \times \int_{x_{0+}}^{b_+} {\rm d} y_+ \int_{y_+}^{b_+} {\rm d} a_+ \int_{y_+}^{x_+} {\rm d} b_+ \int_{b_+}^{x_+} {\rm d} z_+ \int {\rm d}^2 {\bf \rho}_{1 \perp} {\rm d}^2 {\bf \rho}_{2 \perp} {\rm d}^2 {\bf \rho}_{4 \perp} \\
& \quad \times \exp \bigg\{- \int_{x_{0+} }^{y_+} {\rm d} \xi_1 \, \Sigma (\xi_1, x \, {\bf \rho}_{1 \perp}) - \int_{a_+}^{b_+} {\rm d} \xi_3 \, \Sigma (\xi_3, x \, {\bf \rho}_{2 \perp}) - \int_{z_+}^{x_+} {\rm d} \xi_5 \, \Sigma (\xi_5, y \, {\bf \rho}_{4 \perp}) \bigg\} \\
& \quad \times \exp \bigg\{- i \, y \, \bigg({\bf p}_{f \perp} - \frac{1 - x}{x} \, {\bf k}_{1 \perp} - \frac{1 - y}{y} \, {\bf k}_{2 \perp} \bigg) \cdot {\bf \rho}_{4 \perp} - i \, x \, \bigg(\frac{{\bf k}_{1 \perp}}{x} + \frac{{\bf k}_{2 \perp}}{y} \bigg) \cdot {\bf \rho}_{1 \perp} \bigg\} \\
& \quad \times \exp \bigg\{i \, \bigg({\bar q}_1 - \frac{{\bf k}_{1 \perp} \cdot {\bf k}_{2 \perp}}{y \, (1 - x) \, p_+} \bigg) \, (y_+ - a_+) \bigg\} \exp \bigg\{i \, \bigg({\bar q}_2 - \frac{{\bf k}_{1 \perp} \cdot {\bf k}_{2 \perp}}{x \, p_+} \bigg) \, (b_+ - z_+) \bigg\} \\
& \quad \times \bigg(\frac{\partial}{\partial {\bf \rho}_{1 \perp}} - i \, \frac{x}{y} \, {\bf k}_{2 \perp} \bigg) \cdot \bigg(\frac{\partial}{\partial {\bf \rho}_{2 \perp}} + i \, \frac{x}{y} \, {\bf k}_{2 \perp} \bigg) {\cal K}({\bf \rho}_{1 \perp}, y_+; {\bf \rho}_{2 \perp}, a_+ |\mu_1) \\
& \quad \times \bigg(\frac{\partial}{\partial {\bf \rho}_{2 \perp}} - i \, (1 - x) \, {\bf k}_{1 \perp} \bigg) \cdot \bigg(\frac{\partial}{\partial {\bf \rho}_{4 \perp}} + i \, \frac{(1 - x) \, y}{x} \, {\bf k}_{1 \perp} \bigg) {\cal K} \bigg(\frac{x}{y} \, {\bf \rho}_{2 \perp}, b_+; {\bf \rho}_{4 \perp}, z_+ |\mu_3 \bigg),
\end{split}
\end{equation}
where ${\bar q}_1=\frac{x \, m_q^2}{2 \, (1 - x) \, p_+}$ and ${\bar q}_2=\frac{y \, m_q^2}{2 \, (1 - y) \, (1 - x) \, p_+}$ are the reciprocals of the photon formation lengths, $m_q$ is the mass of the quark and $\Sigma (\xi_1, x \, {\bf \rho}_{1 \perp})=n(\xi_1) \sigma(x \, {\bf \rho}_{1 \perp})/2$ is the product of the elastic Mott cross section and the density of scattering centers in the medium. Finally, $\mu_1=(1 - x) x \, p_+$ and $\mu_3=(1 - y) y (1 - x) p_+$. The path integrals are given by
\begin{equation}
{\cal K}({\bf \rho}_{1 \perp}, y_+; {\bf \rho}_{2 \perp}, x_+ |\mu_1) = \int {\cal D} {\bf r}_{ \perp} \exp \left\{\int_{y_+}^{x_+} \!\!{\rm d} \xi_2 \left[\frac{i \, \mu_1}{2} \, {\dot {\bf r}}_{\perp}^2 - \Sigma (\xi_2, x \, {\bf r}_{\perp})\right] \right\} \,,
\end{equation}
where ${\bf r}_\perp(y_+) = {\bf \rho}_{1\perp}$ and ${\bf r}_\perp(x_+) = {\bf \rho}_{2\perp}$ are the boundary conditions. In the dipole approximation, where $\sigma(x \, {\bf \rho}_{1 \perp})\approx C x^2 {\bf \rho}_{1 \perp}^2$, the path integral is given as a solution to the harmonic oscillator with imaginary frequency $\Omega_1=(1 - i)\sqrt{ n_0 \, C \, x^2/(2 \, \mu_1)}$ \cite{WiedemannGyulassy}, namely
\begin{equation}
\begin{split}
& {\cal K}_{osc}({\bf \rho}_{1 \perp}, y_+; {\bf \rho}_{2 \perp}, a_+ |\mu_1) = \frac{\mu_1 \, \Omega_1}{2 \, \pi \, i \, \sin (\Omega_1 \, (a_+ - y_+))} \\
& \quad \times \exp \bigg\{\frac{i \, \mu_1 \, \Omega_1 \, [({\bf \rho}_{1 \perp}^2 + {\bf \rho}_{2 \perp}^2) \cos (\Omega_1 \, (a_+ - y_+)) - 2 \, {\bf \rho}_{1 \perp} \cdot {\bf \rho}_{2 \perp}]}{2 \, \sin (\Omega_1 \, (a_+ - y_+))} \bigg\} \;.
\end{split}
\end{equation}
In this case, and furthermore in the Moli\`ere limit, one can perform all integrations in eq.~(\ref{eq:LadderCrossSection}) analytically. Thus, the spectrum reads
\begin{equation}
\label{eq:TwoPhotonSpectrumMoliere}
\frac{{\rm d}^8 I^\text{Mol}}{{\rm d}^2 {\bf p}_{f \perp} {\rm d}^2 {\bf k}_{1 \perp} {\rm d}^2 {\bf k}_{2 \perp} {\rm d} (\ln x) {\rm d} (\ln y)} \propto \exp \bigg\{- \frac{({\bf k}_{1 \perp} + {\bf k}_{2 \perp} + {\bf p}_{f \perp})^2}{2 \, n_0 \, C \, L_+} \bigg\},
\end{equation}
where $|{\bf k}_{1 \perp} + {\bf k}_{2 \perp} + {\bf p}_{f \perp}|$ $=$ $\sum_i {\bf q}_{i \perp}$ is the total momentum transfer from the medium. One can draw several important conclusions from this, seemingly simple, final result. The appearance of the Moli\`ere factor in eq.~(\ref{eq:TwoPhotonSpectrumMoliere}) signals that the LPM effect still holds for the medium induced two-photon ladder emission. In particular, eq.~(\ref{eq:TwoPhotonSpectrumMoliere}) is not simply a superposition of two one-photon spectra, as one naively would have expected from independent radiation, e.g., in vacuum. Thus, the medium induces some coherence on the multi-emission spectrum. The details of these coherence effects are still under investigation.

\section{Summary and outlook}
We present the study of the spectrum of one- and two-photon radiation off a quark in medium. In the Moli\`ere limit, we find similar features for both, namely the appearance of Gaussian distribution of the total momentum transfer from the medium. In the latter case, this is quite surprising as the corresponding vacuum spectrum is a superposition of two independent emissions.

The above conclusions hold, strictly speaking, only for the ladder emission diagram, depicted in Figure~\ref{twophotonladder1}. For the two-photon case, there appears Feynman diagrams with much more complicated emission sequences, in addition to the diagrams including interference with vacuum emissions. We are still working on a complete, analytical solution of all diagrams.

\end{document}